\documentclass[prl,reprint,10pt,aps,longbibliography,groupedaddress]{revtex4-1}

\usepackage{bbm}
\usepackage{graphicx}
\usepackage{subfigure}
\usepackage{amsmath}

\begin{document}

\title{Directing entanglement spreading by means of a quantum East/West heterojunction structure}

\author{Guanhua Chen$^{1}$ and Yao Yao$^{1,2}$}\email{yaoyao2016@scut.edu.cn}

\address{$^1$ Department of Physics, South China University of Technology, Guangzhou 510640, China\\
$^2$ State Key Laboratory of Luminescent Materials and Devices, South China University of Technology, Guangzhou 510640, China}

\date{\today}

\begin{abstract}
We extend the translationally invariant quantum East model to an inhomogeneous chain with East/West heterojunction structure. In analogy to the quantum diffusion of substantial particles, we observe a cyclic entanglement entropy spreading in the heterojunction during time evolution, which can be regarded as continuous cycles in a quantum heat engine. In order to figure out the possibility of manipulating the entanglement entropy as a quantum resource, the entropy growth is shown to be determined by the initial occupation and the site-dependent chemical potential, and the former is equivalent to an effective temperature. Through fine adjustment of these parameters, we discover the entanglement flow is simply superposed with those from two sources of the chain. An intriguing relation between our model and the traditional heat engines is subsequently established.
\end{abstract}

\maketitle

$Introduction.$ The quantum entanglement, an inherent parameter with nonlocality, has generated much interest as an important physical resource for quantum information processing, metrology and communication \cite{2009rmpent,2010natmetrology,2011natpho,2014jpa,2019rpp,2019rmpres}, which provides a novel perspective to study complicated kinetic and dynamic issues. In recent studies, it was found that the subsystems entanglement entropy as the order parameter in thermalized systems tended to grow rapidly to saturate, while the propagation of correlation could be confined in localized systems and violated the eigenstate thermalization hypothesis, showing a more sluggish logarithmic entropy growth \cite{2012prlentropy,2013prlentropy,2017otoc}. Additionally, in Rydberg-blockaded atomic lattices, the dynamical simulation of irreversible half-chain von Neumman entropy manifests an oscillating increasing tendency, which is named as quantum many-body scars due to strong overlap between ``scarred" eigenstates and specific initial states \cite{2017natscar,2018npscar,2018prbscar,2021npscar}. In one dimensional random quantum circuits, the growth of entanglement is explicitly analogy with stochastic surface growth model of classical particles and the hydrodynamics can be described by the Kardar-Parisi-Zhang equation \cite{2017prxkpz,2020eplkpz,2023annrevkpz}. In this context, it is intuitive to consider feasible ways to efficiently manipulate entanglement entropy in nonthermal systems.

In classical thermodynamics, one is able to manipulate the heat flow by embedding the system into hot and cold baths, but it is difficult to explicitly quantify and direct the relevant entropy flow. In order to extend the classical heat engines to their quantum counterpart, many achievable quantum devices have been studied, such as the thermal diode, thermal transistor and thermal valve \cite{2018prbdiode,2016prltransistor,2012nanovalve,2014annualengine,2019prlengine}. One question then arises is whether these thermal operations can be formulated in the framework of resource theory for entanglement entropy. The subject of this letter is thus to control the flow of entanglement entropy in a kinetically constrained model (KCM) similar with that for conventional heat current and quantitatively direct the entanglement flows by defining the analogical chemical potential and temperature.

$Model.$ We construct a heterojunction based on KCM for the spin-1/2 unidirectional facilitation, e.g., the quantum East model which shows a disorder-free slow dynamical phase transition \cite{2015prbeast,2019prleast,2020prxeast}. The model Hamiltonian is written as
\begin{equation}\label{eq1}
	H_{\rm east}=\frac{1}{2}\sum_{i=1}^{N-1}\mu_in_i-\frac{1}{2}\sum_{i=1}^{N-1}n_i\sigma^x_{i+1},
\end{equation}
where $n_i=(\mathbbm{1}-\sigma^z_i)/2=|1\rangle\langle1|_i$ and $\sigma^{x,z}_i$ are the projector onto the occupied state $|1\rangle$ and Pauli operators at site $i$. Herein, we use the chemical potential $\mu_i$ to replace the common manipulation parameter $e^{-s}$ of thermalization, which is equivalent to the original Hamiltonian \cite{2015prbeast}.

Due to the non-commutation of the two terms in the Hamiltonian, there are not conserved particle numbers, so the fast ($0<\mu_i<1$) or slow ($\mu_i>1$) dynamic phase could be observed on two sides of the Rokhsar-Kivelson (RK) point, respectively, suggesting the chemical potential acts as an energy barrier in the chain. In the last few years, for the quantum East model, eigenstate localization properties and the non-ergodic phase transition under periodic driving have been studied in great details \cite{2020prxeast,2022prxqeast,2022jpbeast,2021prl2deast,2023prl2deast}. As a benchmark, this unidirectional constraint has been extended in higher-dimensional quantum North-or-East model and bosonic quantum East model which has potential to be experimentally implemented in superconducting circuits.

In order to construct the chain with heterojunction structure, we add an analogical quantum West model in contact from the right to the chain and a hopping junction is set between them. The Hamiltonian becomes
\begin{equation}\label{eq2}
	H=H_{\rm east}+H_{\rm cont}+H_{\rm west},
\end{equation}
with
$$
	H_{\rm cont}=-\frac{1}{2}(\sigma^+_{N}\sigma^-_{N+1}+\sigma^+_{N}\sigma^-_{N+1}),
$$
$$
H_{\rm west}=\frac{1}{2}\sum^{2N-1}_{i=N+1}\mu_{i+1}n_{i+1}-\frac{1}{2}\sum^{2N-1}_{i=N+1}\sigma^x_{i}n_{i+1}.
$$
It can be found that the West model has a similar structure with that of the East model but with an opposite facilitation direction. The middle two sites are symmetrically set as a contact which are unable to influence other sites but can be controlled by them.

\begin{figure}[htbp]
	\includegraphics[scale=0.9]{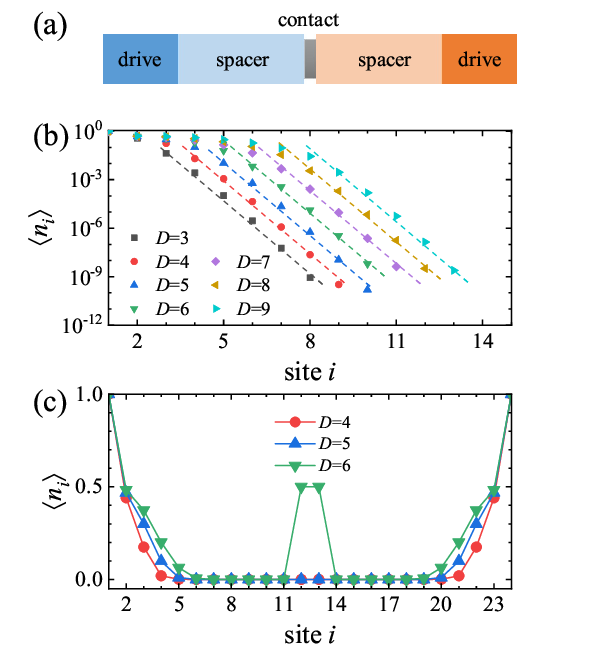}
	\caption{\label{fig1} (a) Schematic illustration of our model with heterojunction structure. East and West parts are marked in blue and orange, respectively. Drive and spacer regions are distinguished by differences in the shade of color. Middle contact interaction is ploted as gray. (b) The simulated mean occupation $\langle n_i\rangle$ of ground states of quantum East model ($N=24$, only partial sites are shown) with seven different $D$'s. Exponential decay is observed through dashed lines of corresponding color. (c) Ground-state $\langle n_i\rangle$ of the heterojunction chain with $D=4$ (red), $D=5$ (blue) and $D=6$ (green).}
\end{figure}

For the sake of generation of current flow, we introduce two regions on each side of the contact: The spacer region with chemical potentials of all sites being set to $\mu_i=2$ throughout this paper, and the drive region, unless particularly stated, with chemical potentials being set as $\mu_i=0.99$, as depicted schematically in Fig.~\ref{fig1}(a). Notice that the chemical potential on $i$-th site actually controls $(i+1)$-th site, so the last drive site is also set to $\mu_i=2$ and we define the number of drive sites $D$ as the drive region size. On the other hand, the number of occupied state $|1\rangle$ is essential which serves as the source to facilitate neighboring sites. Since the number of $|1\rangle$ is not conserved, the number of initial $|1\rangle$ states can be regarded as an ``effective temperature", as discussed below. By this consideration, we initially set all the spacer sites to be $|0\rangle$ (namely zero temperature) and initial states on the drive sites can be changed within $|0\rangle$ and $|1\rangle$. We define initial occupation number $\rho_0$ to quantify the number of $|1\rangle$ in drive regions. For example, a chain with $\mu_1=\mu_2=0.99$, $\mu_3=\dots=\mu_{2N-1}=2$ implies $D=3$ and the 8-site state $|101000\dots00\rangle$ has initial occupation $\rho_0=2$. Different from the translation-invariant feature of the East model, both the site-dependent $\mu_i$ and $\rho_0$ provide variables to manipulate the imbalanced particle densities flowing. The most advantage of this model is that, since the total number of particle is not conserved, drive regions with initial $|1\rangle$ states can be regarded as sources to generate the steady flow, which is unavailable in conserved systems.

$Results.$ In the following, we numerically investigate the ground states and dynamics with default size $L=2N=24$ by density matrix renormalization group (DMRG) and time evolving block decimation (TEBD) methods in open boundary condition (OBC) \cite{2007tebd,2011dmrg,ITensor}. Based on previous researches, the ground state of translation-invariant quantum East model for $\mu_i>1$ shows that occupied sites are localized at the first few sites. That is, the expected value of $n_i$ exponentially decays as $\langle n_i\rangle \sim e^{-i/\xi}$, with the localization length $\xi$ depending on $\mu_i$ \cite{2020prxeast}. By contrast, a homogeneous occupation distribution is present in thermal phase ($\mu_i<1$ or $s<0$). By adding non-uniform $\mu_i$ to the quantum East model, in Fig.~\ref{fig1}(b), we show the mean occupation $\langle n_i\rangle$ as a function of sites $i$ with different $D$. Different from that in the previous researches, the fitting dotted lines which figure out the decay rates have exactly the same slope. On the other hand as illustrated by Fig.~\ref{fig1}(c), by changing the number of drive sites up to $D=6$ in the heterojunction model, the particles are found to accumulate in the middle of the chain. It implies that enlarged drive region can not only increases the occupation in the drive region but also enforces the particles injecting onto the surface of spacer region, similar with the function of gate voltage in conventional field-effect transistors.

\begin{figure}[htbp]
	\includegraphics[scale=0.9]{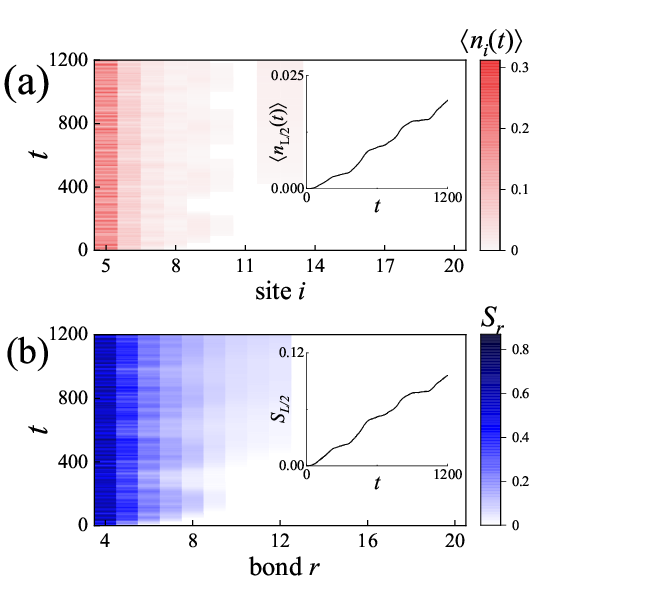}
	\caption{\label{fig2} Dynamics of initial state as $|11100\dots 000\rangle$ for $D=3$. (a) Time evolution of occupation $\langle n_i\rangle$.  The growth of middle site $\langle n_{L/2}\rangle$ is specifically shown in the inset. (b) The von Neumann entropy $S_r$ of every bond $r$, i.e. the subsystem of reduced density matrix is chosen based on the cut between $i=r$ and $i=r+1$. The bipartition entanglement evolution of middle bond $S_{L/2}$ is present in the inset. }
\end{figure}

For investigating this injection of particles in a dynamic manner, we now construct a unipolar East model and then extend to more complicated structures later on. As mentioned in Ref.\cite{2020prxeast}, the Hilbert space of quantum East model is determined by the first occupied site which is normally set as the leftmost sites and vice verse for West model, so the choice of initial state is particularly important. Here, we show the unipolar evolution of the initial state $\rho_0=3$, $D=3$, namely $|111000\dots00\rangle$. As shown in Fig.~\ref{fig2}(a), the dynamics of $\langle n_i\rangle$ performs a predictable particles accumulation at the leftmost site, manifesting a localized slow diffusion process.

As argued in recent studies, the growth of entanglement entropy holds much insightful physics, such as the measurement-induced phase transitions in which the measurement-rate-dependent transitions from volume-law to area-law have been observed \cite{2019measurement,2020prrmeasure,2023prlmeasure,2023npmeasure}. The bond-dependent entanglement entropy turns out to be essential in studying this phase transition, as well as the quantum Zeno effect\cite{2018zeno}. We thus explore relevant entanglement spreading by considering the entropy cut at every bond of the chain, namely $S_r=-{\rm Tr}_r\varrho{\rm log}\varrho$, with $\varrho$ being the reduced density matrix taken partial trace at $i=r$, which is different from the frequently-used half-chain von Neumann entanglement entropy. In Fig.~\ref{fig2}(b), the entanglement entropy manifests similar evolution features with that of mean occupations, that is, there are four periods of oscillation up to around $t\sim350$. In order to be more detailed on these similar variation tendencies, we plot the $\langle n_{L/2}\rangle$ and half-chain entropy $S_{L/2}$ in the insets of Fig.~\ref{fig2}. It is clear that they share almost the same growth process, resulting from the blocking effect of $H_{\rm cont}$ on the entanglement spreading.

The wave-like entanglement spreading from sources accumulate at junction with rate changing periodically, generating a cyclic entropy growth, which means the crests are associated with maximums of growth rate and every trough resulting in the slowest growth corresponds to quasi-isentropic processes. This oscillatory behavior of entanglement entropy can be regarded as continuous cycles of heat engine. Furthermore, since the evolution on the sites from $i=L/2+2$ to $L$ can be ignored, the reduced density matrix after partial tracing the left half chain is totally determined by the site $i=L/2+1$. That is, in terms of the property of entanglement entropy, $S_{L/2}$ can only depend on $\langle n_{L/2+1}\rangle$ (and also $\langle n_{L/2}\rangle$ in terms of the hopping junction). This entanglement growth analogous to particles reminds us of the surface growth model for entanglement in random quantum circuits, where an entanglement tsunami can be described by the KPZ equation of particles \cite{2010prlkpz,2014prltsunami,2017prxkpz}. We further calculate the entropy growth of systems of different sizes which solely shows the entanglement of smaller system is easier to approach middle junction.

\begin{figure}[htbp]
	\includegraphics[scale=1.0]{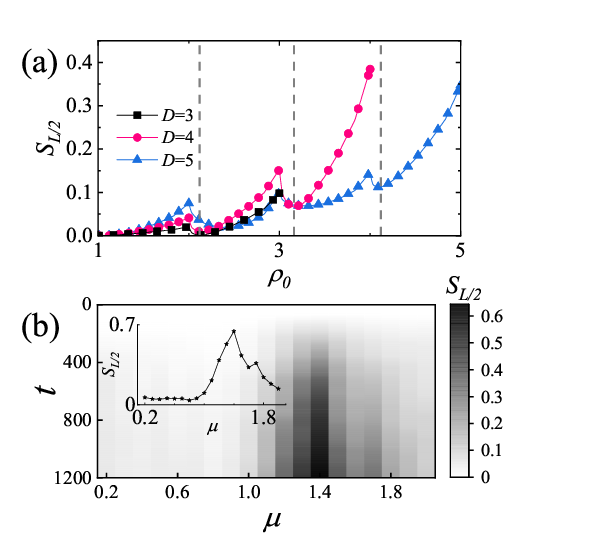}
	\caption{\label{fig3} The entanglement entropy under different conditions in the heterojunction. (a) The $S_{L/2}$ at $t=1200$ as a function of the initial occupation $\rho_0$ with $D=3$ (black square), $D=4$ (pink circle) and $D=5$ (blue triangle). Three local minimal values are marked with dashed lines. (b) The evolution of $S_{L/2}$ as a function of $\mu_1=\mu_2=\mu$. Inset: The final states $S_{L/2}$ at $t=1200$. }
\end{figure}

Above results imply that we may be able to manipulate entanglement analogous to enabling the particle flow with temperature gradient. Because of the nearest neighbor one-spin facilitation of $H_{\rm east}$, an occupied site may inhibit the next-nearest-neighbour flip. For example, suppose a configuration in three sites of a chain, the evolution results in
\begin{equation}\label{eq4}
	\begin{aligned}
		&&| 1_{1}\rangle(\mu|1\rangle-|0\rangle)_2|0_{3}\rangle, \\
		&\nearrow&& \\
		|1_{1}1_{2}0_{3}\rangle& \\
		&\searrow&& \\
		&&| 1_{1}1_{2}\rangle(-|1\rangle+\mu|0\rangle)_3 ,
	\end{aligned}
\end{equation}
and
\begin{equation}\label{eq5}
	|0_{1}1_{2}0_{3}\rangle\rightarrow|0_{1}1_{2}\rangle(-|1\rangle+\mu|0\rangle)_3,
\end{equation}
where $\rightarrow$ figures out subsequent evolving states. In the former case, the first site indirectly inhibits the facilitation of the third site by flipping the second site, which is not the case in the latter. As a result, an occupied site is able to facilitate the nearest neighbour as well as possibly reduce facilitation of further sites. The initial occupation therefore turns out to be essential, which as stated plays the role of a high-temperature source in the drive region.

To this end, we then focus on the influence of initial occupation and drive size on $S_{L/2}$, which has been discussed a bit based on the maximal number of consecutive down spins in Ref.\cite{2015prbeast}. We construct a series of non-integral $\rho_0$ for initial states as
\begin{equation}\label{eq6}
	|\phi (x,\alpha) \rangle=|1_{1}1_{2}\dots1_{x}\rangle |\alpha\rangle_{x+1}|0_{x+2}\dots 0_{N-1}0_{N}\rangle,
\end{equation}
where the superposition state is set as $|\alpha\rangle=\sqrt{\alpha}|1\rangle+\sqrt{1-\alpha}|0\rangle$ and the initial occupation becomes $\rho_0=x+\alpha$. Take an example: $|\phi(3,1/3)\rangle=|111\rangle(\sqrt{1/3}|1\rangle+\sqrt{2/3}|0\rangle)|000\dots\rangle$. Fig.~\ref{fig3}(a) illustrates $S_{L/2}$ versus $\rho_0$ at $t=1200$ with different $D$. It is observed that every local maximum appears at integral $\rho_0$. More interestingly, as the initial occupation increasing, $S_{L/2}$ shows a small downward tendency and will reach local minimums (marked with dashed lines), which are balanced points between inhibition and facilitation. It is worth stressing that the drive size $D$ does almost not affect the position of minimum. As mentioned above, the $\rho_0$ could be regarded as an effective temperature compared to the ``cold'' spacer region, so Fig.~\ref{fig3}(a) is nothing but an entropy-temperature diagram for the heat engine which is approximately constituted by isentropic and isothermal processes. The former is from the pure state $|\alpha\rangle$ to $|1\rangle$ in a single site without entropy change and the latter is to flip the neighboring site from $|0\rangle$ to $|1\rangle$ enabling entropy increase. This then serves as the essential point of our model as we find a novel way to define a temperature gradient in a quantum heat engine.

One may ask if the chemical potential $\mu_i$ has got the same effect of temperature. We then fix the $D=3$, $\rho_0=3$ and show the entropy variance as a function of $\mu_i$ in drive region in Fig.~\ref{fig3}(b). The entanglement does not change with $\mu$ when $\mu<1.0$ and then increases when $1.0<\mu<1.4$. Crossing a turning point, the smaller $\mu$ enables more active facilitation and more easily spreading entanglement. As addressed above, a possible phase transition is supposed to appear at the critical point $\mu=1$ for fast and slow dynamics, but the fastest-growing entropy as well as the turning point is observed around $\mu=1.4$. This unexpected result can be interpreted as follows. Just like the block layer in field-effect transistors, the spacer region here acts as a block to make the entanglement merely oscillate in the drive region, and then the higher oscillation frequency from lower $\mu$ makes particles difficult to be injected into the spacer region. After increasing $\mu$ greater than the critical point $\mu=1$, the slowdown allows partial entanglement entering the spacer region. At the emergent crossover $\mu=1.4$, the balance between the oscillation and the slow dynamics is reached and subsequently results in the later descent process. In consequence, the chemical potential in our model can not be equivalently recognized as the temperature; it is more like a parameter of detuning.

\begin{figure}[htbp]
	\includegraphics[scale=1.0]{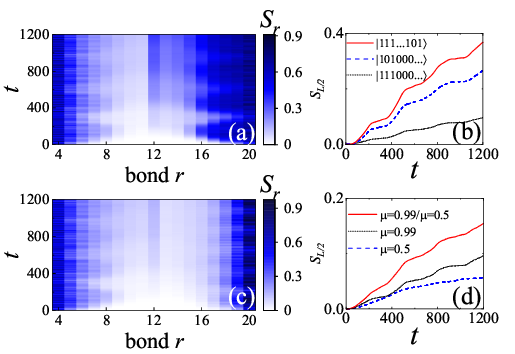}
	\caption{\label{fig4} Dynamics of entanglement entropy $S_r$ for the heterojunction chain. (a) The initial state is chosen as $|11100\dots00101 \rangle$ with $D=3$ and $\mu=0.99$. (c) The initial state is chosen as $|11100\dots00111\rangle$ with $D=3$, $\mu=0.99$ in the East model and $\mu=0.5$ in the West model. Comparisons of $S_{L/2}$ under different conditions are plotted in (b) and (d).}
\end{figure}

Hence, now we have two parameters to regulate the growth of entanglement flows. As seen in Fig.~\ref{fig2}, the entropy flow shows apparent periodic waves whose amplitude and period are related to $\rho_0$ and $\mu_i$. Concretely, we would like to show two examples of entanglement manipulation, couplings between the same periods with different amplitudes and the similar amplitudes with different periods. After some tentative simulations, we have elaborately chosen initial conditions appropriate for analysis. We set the initial state as $|11100\dots00101\rangle$ and plot the resulted $S_r$ in Fig.~\ref{fig4}(a). Even though two waves have the same period from equal $\mu$, they have unequal amplitudes and the West part shows an evidently faster entanglement flow than that in the East part. This amplitude difference could be explained by Eq.(\ref{eq4}) and Eq.(\ref{eq5}). The $S_{L/2}$ of three different initial states as a bipolar and two unipolar evolutions are further displayed in Fig.~\ref{fig4}(b). Three curves have similar cyclic growth tendencies and interestingly the red curve is equivalent to superposition of the other two, which is qualitatively attributed to the extensive property of entropy. In addition, we calculate the evolution under two different dynamical parameters, $\mu=0.99$ in the East and $\mu=0.5$ in the West as shown in Fig.~\ref{fig4}(c). Here, a lower $\mu$ brings lower period, that is a coupling between two different periods and similar amplitudes. A simple superposition is also found in Fig.~\ref{fig4}(d).

$Discussion\ and\ conclusion.$ Before ending, we discuss more on the chemical potential and temperature. In our model, we replace the flip factor $e^{-s}$ in the common quantum East model with parameter $\mu_i$ as the chemical potential for the occupation $n_i$ in Hamiltonian Eq.(\ref{eq1}). As well known, in the traditional thermodynamics, the internal energy of ideal gases can be described by the Euler equation as
\begin{equation}\label{eq7}
	E=TS-PV+\sum_j\mu_j N_j,
\end{equation}
where $N_j$ denotes the particle number of $j$-th component and $\mu_j$ the relevant chemical potential. In an electrochemical cell, the difference of chemical potential serves as the bias voltage to generate current flow. In the quantum East model, however, the chemical potential difference not only facilitates but also inhibits the spreading of particles as well as the entanglement entropy. This dualistic role of chemical potential mainly stems from that the particle number $n_i$ and entanglement entropy $S_r$ have almost identical growth process, which suggests a novel way to understand the relationship between classical and quantum heat engine. Most importantly, since in normal cases the latter is based upon pure states, it is difficult to define a relevant temperature analogous to its classical counterpart. Here in our model, the drive region size and initial occupation together play the role of temperature, which can be promisingly applied to redefine the temperature in quantum heat engines.

In summary, we have studied a non-uniform 1D KCM based on the nearest-neighbor facilitated quantum East model. We focus on the dynamics of entanglement spreading and further explore its potential of realizing entanglement manipulation. The entanglement flows depending on the initial states and drive region are investigated. Through introducing an analogical West model to construct a heterojunction, the superposed entanglement flows are studied and the relationship with quantum heat engines is discussed.

\section{Acknowledgments}

The authors gratefully acknowledge support from the National Natural Science Foundation of China (Grant Nos.~12374107 and 11974118).

\bibliography{ref}
\end{document}